\documentclass[12pt,letter]{article}
\usepackage[left=1.3in,top=1.25in,right=1.3in,bottom=1.25in]{geometry}
\usepackage{amsfonts}
\usepackage{amssymb}
\usepackage{graphicx}
\usepackage{setspace}
\usepackage[round]{natbib}
\usepackage{amsmath}
\usepackage{amssymb}
\usepackage{amsbsy}
\usepackage{amsthm}
\usepackage{paralist}

\newtheorem{theorem}{Theorem}

\newtheorem{definition}{Definition}

\newtheorem{lemma}[theorem]{Lemma}

\newcommand{\bX}{\bf{X}}
\newcommand{\Oi}{\Omega^{(g)}}
\newcommand{\bone}{\bf{1}}
\def\E{\qopname\relax o{E}}
\newcommand{\diag}{\mbox{diag}}
\newcommand{\tr}{\mbox{tr}}

\newcommand{\N}{\mathcal{N}}
\newcommand{\PG}{\mathcal{PG}}
\newcommand{\Ga}{\mbox{Ga}}
\newcommand{\Pol}{\mbox{Pol}}

\title{Default Bayesian analysis for multi-way tables: \\ a data-augmentation approach}
\author{Nicholas G Polson\\
\textit{University of Chicago}\footnote{Polson is Professor of Econometrics and Statistics
at the Chicago Booth School of Business. email: ngp@chicagobooth.edu. Scott
is Assistant Professor of Statistics at the University of Texas at Austin.
email: James.Scott@mccombs.utexas.edu.}\\ \\
James G. Scott\\
\textit{University of Texas at Austin}}
\date{First Draft: August 2011\\
This Draft: September 2011}

\begin{document}

\maketitle
\begin{abstract}
\noindent This paper proposes a strategy for regularized estimation in multi-way contingency tables, which are common in meta-analyses and multi-center clinical trials.  Our approach is based on data augmentation, and appeals heavily to a novel class of Polya--Gamma distributions.  Our main contributions are to build up the relevant distributional theory and to demonstrate three useful features of this data-augmentation scheme.  First, it leads to simple EM and Gibbs-sampling algorithms for posterior inference, circumventing the need for analytic approximations, numerical integration, Metropolis--Hastings, or variational methods.  Second, it allows modelers much more flexibility when choosing priors, which have traditionally come from the Dirichlet or logistic-normal family.  For example, our approach allows users to incorporate Bayesian analogues of classical penalized-likelihood techniques (e.g.~the lasso or bridge) in computing regularized estimates for log-odds ratios.  Finally, our data-augmentation scheme naturally suggests a default strategy for prior selection based on the logistic-Z model, which is strongly related to Jeffreys' prior for a binomial proportion.  To illustrate the method we focus primarily on the particular case of a meta-analysis/multi-center study (or a $J \times K \times N$ table).  But the general approach encompasses many other common situations, of which we will provide examples.
\end{abstract}

\newpage

\begin{spacing}{1.05}

\section{Introduction}

In this paper, we fit hierarchical Bayesian models for multi-way contingency tables using data augmentation.  We focus on $J\times K \times N$ tables, which are common in meta-analyses or multi-center studies.  One reason for the relative dearth of practical, exact Bayesian approaches to these problems is the nonlinearity (and associated nonconjugacy) of their likelihoods.  Our data-augmentation approach directly addresses this issue.  It thereby avoids the need for analytic approximations, numerical integration, Metropolis--Hastings, or variational methods.  We also describe many extensions involving logistic-type models that rely upon the same basic framework.  These extensions encompass many areas of wide interest in modern statistical practice, including mixtures of logits and mixed-membership/topic models.

Table 1 presents a simple example of a multi-way table, from \citet{skene:wakefield:1990}.  The data arise from a multi-center trial comparing the efficacy of two treatment arms---in this case, different topical cream preparations, labeled the treatment and the control.  This table suggests two advantages of pooling information across treatment centers in a hierarchical Bayesian model, both articulated by many previous authors: it sharpens the estimate of the overall difference between treatment and control, and it regularizes those proportions whose maximum-likelihood estimates would otherwise be identically zero---for example, in the control group at centers 5 and 6, where no successes were observed.

The goal of this paper is not to propose new statistical models for such tables.  Rather, we work firmly within the context of existing models, showing:
\begin{inparaenum}[(1)]
\item that these models have a normal mixture representation involving Polya--Gamma random variables;
\item that this mixture representation is of great practical relevance, since it leads to new, efficient Gibbs and EM algorithms for posterior computation; and
\item that this representation also allows users far greater flexibility in specifying computationally tractable priors, including default or ``objective'' priors.
\end{inparaenum}

Our fundamental contribution is the data-augmentation scheme of Section \ref{sec:PGaugmentation}, which appeals heavily to a novel class of Polya--Gamma distributions.  The associated distributional theory depends, in turn, upon the L\'evy representation of Fisher's Z distribution, discussed extensively by \citet{Polson:Scott:2010b}.  This allows us to represent logistic likelihoods directly as mixtures of normals.  The resulting mixture representation is quite parsimonious, in that it involves only one latent variable per cell in the table.  This compares favorably with other, different latent-variable methods that involve one latent variable per observation---for example, the random-utility representation of the logit model used by \citet{holmes:held:2006} in the context of logistic regression.

Our work is closest to that of \citet{leonard:1975}, \citet{skene:wakefield:1990}, and \citet{forster:skene:1994} in two respects: we parameterize tables in terms of log-odds ratios; and we use logistic-normal priors, along with their generalizations, within a hierarchical Bayesian model. It is also closely related with the work of \citet{gelman:etal:2008} and \citet{bedrick:christensen:johnson:1996}, and to a lesser extent that of \citet{gelman:2006}, in that we propose a new default prior for log-odds ratios in logistic-type models.

Extensive bibliographies on Bayesian methods for categorical data analysis can be found in \citet{agresti:hitchcock:2005}, \citet{forster:2011}, and Chapter 5.7 of \citet{gelman:carlin:stern:rubin:2004}.  Similar issues of computational tractability arise in ecological inference for $2 \times 2$ tables \citep{wakefield:2004}; in other models for meta-analysis \citep{gray:1994,carlin:1992}; and in Poisson-type models for $2 \times 2 \times N$ tables \citep{demir:hamur:2008}.  Other important Bayesian papers on contingency tables include \citet{altham:1969} and \citet{crook:good:1980}.  While the focus of our analysis is not on testing for independence, in the manner of \citet{diaconis:efron:1985}, it is also possible to compute Bayes factors using our algorithms.

\begin{table}
\caption{\label{tab:skenewakeexample} Data from a multi-center, binary-response study on topical cream effectiveness \citep{skene:wakefield:1990}.}
\vspace{1pc}
\begin{center}
\begin{tabular}{c  cccc}
\hline
 & \multicolumn{2}{c}{Treatment} & \multicolumn{2}{c}{Control} \\
Center & Success & Total & Success & Total \\\hline
1 & 11 & 36 & 10 & 37\\
2 & 16 & 20 & 22 & 32\\
3 & 14 & 19 & 7 & 19\\
4 & 2 & 16 & 1 & 17\\
5 & 6 & 17 & 0 & 12\\
6 & 1 & 11 & 0 & 10\\
7 & 1 & 5 & 1 & 9\\
8 & 4 & 6 & 6 & 7\\
\hline
\end{tabular}
\end{center}
\end{table}

The paper is organized as follows.  Section \ref{sec:PGaugmentation} presents our main result concerning the Polya--Gamma mixture representation for the logit-type likelihoods that arise in multi-way tables, beginning with the $2 \times 2$ case and working up from there.  Sections \ref{sec:PGEM} and \ref{sec:PGGibbs} use this representation to derive simple EM and Gibbs-sampling algorithms, respectively, for posterior computation. Section \ref{sec:skenewakefieldexample} analyzes the data from Table \ref{tab:skenewakeexample} to illustrate the proposed method.  Section \ref{sec:priorchoice} proposes a default logistic-Z prior for the models of Section \ref{sec:PGaugmentation}, appealing to Jeffreys' arguments about priors for binomial proportions.  Section \ref{sec:generalizations} considers the general $J \times K \times N$ case.  Section \ref{sec:discussion} concludes with several remarks about generalizations and interesting features of the overall approach.  Proofs of our main results, along with the distributional theory of Polya--Gamma random variables, are deferred to appendices.

\section{Data augmentation for multi-way tables}

\label{sec:PGaugmentation}

\subsection{A single table}

First, consider the situation of a binary-response trial designed to compare an active treatment with a control.  We will use this simple case to introduce the basic theory behind our approach, before generalizing to the case of a multi-center multinomial response trial (or a $J \times K \times N$ table).

For the treatment group, we observe $y_{1}$ successes among $n_1$ subjects, while for the control group, we observe $y_2$ successes among $n_2$ subjects.  Let $p_1$ denote the underlying success probability for the active treatment, and $p_2$ the success probability for the control.  Let $p = (p_1, p_2)'$.  Clearly the likelihood is a product of binomial probability mass functions: $L(p) = (p_1)^{y_1} (1-p_1)^{n_1 - y_1} \ (p_2)^{y_2} (1-p_2)^{n_2 - y_2}$. 

Let $\psi_1$ and $\psi_2$ denote the log-odds ratios corresponding to $p_1$ and $p_2$, respectively:
$$
\psi_1 = \log \left(  \frac{p_1}{1-p_1} \right) \quad \mbox{and} \quad \psi_2 = \log \left(  \frac{p_2}{1-p_2} \right) \, .
$$
We first consider the case where $\psi$ is assigned a bivariate normal prior with mean $\mu$ and covariance matrix $\Sigma$, both fixed in advance. The posterior distribution given data $D = \{y_1, n_1, y_2, n_2\}$ is
\begin{eqnarray}
p(\psi \mid D) &\propto& p(\psi) \ p(D \mid \psi) \nonumber \\
 &=& \exp\left\{ -\frac{1}{2} (\psi - \mu)'\Sigma^{-1} (\psi-\mu) \right\} 
\frac{( e^{\psi_1})^{y_1} } {( 1 + e^{\psi_1})^{n_1} }   \cdot \frac{( e^{\psi_2})^{y_2} } {( 1 + e^{\psi_2})^{n_2} } \label{eqn:posterior2x2} \, ,
\end{eqnarray}
where we have re-written the likelihood in terms of the log-odds ratios $\psi=(\psi_1, \psi_2)'$.  This does not factorize easily into an analytically convenient form, and has traditionally been analyzed using numerical integration \citep[c.f.][]{skene:wakefield:1990}, analytic approximations to the likelihood \citep{carlin:1992,gelman:carlin:stern:rubin:2004}, or Metropolis-Hastings \citep{dobra:tebaldi:west:2006}.

Our main result is that this posterior distribution, far from being intractable, is actually a mixture of bivariate normal distributions.  So too are many logit-type likelihoods similar in overall structure.  This leads to very simple EM and MCMC schemes for posterior computation; we will present these algorithms shortly, after describing the mixing distribution itself.

The mixing distribution in this conditionally normal representation is from a new class of random variables, which we call the Polya--Gamma class.  As their name suggests, Polya--Gamma distributions are closely related to Polya distributions, or infinite convolutions of exponentials.  In the appendix, we summarize some basic facts about the Polya--Gamma distribution, including its density and moment-generating function.  (A special case of the density, for example, is a multi-scale mixture of inverse-Gaussians, which is an interesting generalization of a standard prior.)  Here, we simply introduce the class in a manner that lends itself straightforwardly to simulation.

\begin{definition}
A random variable $X \sim \PG(a,c)$ has a Polya--Gamma distribution  if
\begin{equation}
\label{eqn:PGdistribution1}
X \stackrel{D}{=} \frac{1}{2 \pi^2} \sum_{k=1}^{\infty} \frac{g_k}{(k-1/2)^2 + c^2/(4\pi^2)} \, ,
\end{equation}
where each $g_k$ is an independent gamma random variable: $g_k \sim \Ga(a,1)$.
\end{definition}

We can now state the main result of the paper, which is proven in the appendix.

\begin{theorem}
\label{thm:2x2PGmixture}
The posterior distribution in (\ref{eqn:posterior2x2}) is a mixture of normals with respect to latent variables $\Omega = \mbox{diag}(\omega_1, \omega_2)$. This mixture takes the following form.

\begin{description}
\item[Part A:] The posterior $p(\psi \mid D)$ can be expressed hierarchically as
\begin{eqnarray}
(\psi \mid D, \Omega) &\sim& \N(m_{\Omega}, V_{\Omega}) \label{eqn:pgmixture1}
\end{eqnarray}
where $\omega_{j}$ are latent variables with prior distribution
\begin{equation}
\label{eqn:pgmixture2}
\omega_j \sim \PG(n_j, 0)  \, ,
\end{equation}
for $j = 1,2$; and where
\begin{eqnarray*}
V_{\Omega}^{-1} &=& \Omega + \Sigma^{-1} \\
m_{\Omega} &=& V(\kappa + \Sigma^{-1} \mu) \\
\kappa &=& (y_1 - n_1/2, y_2 - n_2/2)' \, .
\end{eqnarray*}

\item[Part B:] The conditional posterior $p(\omega_j \mid \psi_j, D)$ arising from the model in (\ref{eqn:pgmixture1})--(\ref{eqn:pgmixture2}) is also in the Polya--Gamma family:
\begin{equation}
(\omega_j \mid \psi_j, D) \sim \PG\left( n_j, \psi_j \right)
\end{equation}
for $j = 1,2$.

\end{description}

\end{theorem}

\subsection{A series of tables}

Suppose now that similar binary-response trials are conducted in each of $N$ different treatment centers.  Let $n_{ij}$ be the number of patients assigned to regime $j$ in center $i$, and let $y_{ij}$ be the corresponding number of successes, for $i=1, \ldots, N$ and for $j=1$ (active treatment) and $j=2$ (control).  As for the case of a single table, let $p_{ij}$ denote the underlying success probabilities, and $\psi_{ij}$ the corresponding log-odds ratios in favor of success.

Assuming that the individual terms $\psi$ are conditionally independent (given some common set of hyperparameters), the posterior for $\Psi = \{\psi_{ij}\}$ is
$$
p(\Psi \mid D) \propto \prod_{i=1}^N \left\{ \frac{  e^{ y_{i1} \psi_{i1} } }{ ( 1 + e^{\psi_{i1}} )^{n_{i1}} }
 \frac{  e^{ y_{i2} \psi_{i2} } }{ ( 1 + e^{\psi_{i2}} )^{n_{i2}} } \ p( \psi_{i1} , \psi_{i2} ) \right\} \, .
$$
Suppose that, as before, we assume a bivariate normal prior: $\psi_i = (\psi_{i1}, \psi_{i2})' \sim \N(\mu, \Sigma)$.  Applying Part A of Theorem \ref{thm:2x2PGmixture} to each term in the posterior, we may introduce augmentation variables $\Omega_i = \mbox{diag}(\omega_{i1}, \omega_{i2})$ to arrive at the following conditional representation:
\begin{eqnarray}
(\psi_i \mid D, \Omega_i) &\sim& \N(m_i, V_{\Omega_i}) \label{eqn:pgmixture3} \\
\omega_{ij} &\sim& \PG(n_{ij}, 0) \quad \mbox{for $j = 1,2$} \nonumber \, ,
\end{eqnarray}
where
\begin{eqnarray*}
V_{\Omega_i}^{-1} &=& \Omega_i + \Sigma^{-1} \\
m_i &=& V_{\Omega_i} (\kappa_i + \Sigma^{-1} \mu) \\
\kappa_i &=& (y_{i1} - n_{i1}/2, y_{i2} - n_{i2}/2)' \, .
\end{eqnarray*}

Moreover, applying Part B,
\begin{equation}
\label{eqn:pgmixture4}
(\omega_{ij} \mid \psi_{ij}, D) \sim \PG\left( n_{ij}, \psi_{ij} \right) \, .
\end{equation}
We now use this representation to derive simple EM and Gibbs-sampling algorithms for estimating the model parameters.

\section{MAP estimation via EM}
\label{sec:PGEM}

We employ an EM algorithm to estimate the posterior mode for $\Psi$, beginning with the case where $\mu$ and $\Sigma$ are pre-specified.

\begin{figure}[t!]
\begin{center}
\fbox{
\begin{minipage}{32pc}
\begin{center}
\textbf{Algorithm 1: EM for $2\times2\times N$ tables \\
(normal prior, fixed $\mu$ and $\Sigma$)}
\end{center}

Begin with an initial guess $\Psi^{(1)}$.

\begin{description}
\item[For] iteration $g = 1, 2, \ldots$

\begin{description}
\item[E Step:] For $i=1:n$ and $j=1:2$, set
$$
\omega_{ij}^{(g)} :=  \frac{n_{ij}}{\psi^{(g)}_{ij}} \tanh(\psi^{(g)}_{ij}/2) \, .
$$

\item[M Step:] For $i=1:n$ set
$$
\psi^{(g+1)} := \left(\Omega_i^{(g)} + \Sigma^{-1} \right)^{-1} \left(\kappa_i + \Sigma^{-1} \mu \right)
$$
for $\Omega_i^{(g)} = \mbox{diag}(\omega_{i1}^{(g)}, \omega_{i2}^{(g)} )$ and $\kappa_i = (y_{i1} - n_{i1}/2, y_{i2} - n_{i2}/2)'$.

\end{description}

\item[End] when the sequence of estimates $\{\Psi^{(1)}, \Psi^{(2)}, \ldots \}$ has converged.

\end{description}

\end{minipage}
}
\vspace{0.5pc}
\caption{\label{algo:EM2x2.1} An EM algorithm for estimating log-odds ratios in a $2\times2 \times N$ table.}
\end{center}
\end{figure}

Let $\Psi^{(g)}$ denote our current estimate of the vector of log-odds ratios.  In the E step, we compute the expected value of the log posterior distribution, given this current guess, marginally over the augmentation variables $\Omega$.  Since the posterior given $\Omega$ is conditionally normal,
\begin{eqnarray*}
Q(\Psi \mid \Psi^{(g)}) &=& \E  \{ \log p(\Psi \mid D,  \Omega) \} \\
&=& \E\left( \sum_{i=1}^N
 \left\{  \kappa_{i1} \psi_{i1} - \frac{\omega_{i1} \psi_{i1}^2}{2} +  \kappa_{i2} \psi_{i2} + \frac{\omega_{i2} \psi^2_{i2}}{2}  - \frac{1}{2}(\psi_i - \mu)'\Sigma^{-1} (\psi_i - \mu) \right\} \right) \\
 &=&  \sum_{i=1}^N
 \left\{  \kappa_{i1} \psi_{i1} - \frac{\hat{\omega}^{(g)}_{i1} \psi_{i1}^2}{2} +  \kappa_{i2} \psi_{i2} + \frac{\hat{\omega}^{(g)}_{i2} \psi^2_{i2}}{2}  - \frac{1}{2}(\psi_i - \mu)'\Sigma^{-1} (\psi_i - \mu) \right\} \, ,
\end{eqnarray*}
where
$$
\hat{\omega}^{(g)}_{ij} = E\left( \omega_{ij} \mid \psi_{ij}^{(g)} \right) \, .
$$
All expectations are under the conditional posterior distribution for $\Omega$, given the current guess $\Psi^{(g)}$.  This final step is justified because the objective function is linear in the $\omega_{ij}$'s, and because these terms are conditionally independent in the posterior, given the $\psi_{ij}$'s.

In the M step, we maximize this as a function of all the $\psi_i$'s jointly to yield the next estimate, $\Psi^{(g+1)}$.  Since the $\psi_i$'s are conditionally independent in the posterior, given $\Omega$, the maximizing value of $\psi_i$ can be trivially computed using standard normal theory as
$$
\psi_i^{(g+1)} = \left(\Omega_i^{(g)} + \Sigma^{-1} \right)^{-1} \left(\kappa_i + \Sigma^{-1} \mu \right) \,.
$$
This is essentially weighted least squares, although it is unusual in the sense that the weights also appear as part of what would normally be construed as the dependent variable in a regression.  An interesting comparison is with the methods for sparse Bayes estimation proposed by \citet{Polson:Scott:2011a}.

To run the algorithm, it is therefore sufficient to know the conditional expected value of the latent precision $\omega_{ij}$ to plug in to the E Step.  The following lemma, proven in the appendix, allows this quantity to be computed without difficulty.
\begin{lemma}
\label{lem:PGmoment}
Suppose $X \sim \PG(a,c)$.  Then
\begin{equation}
\label{eqn:PGmoment}
\E(X) = \frac{a}{c} \tanh(c/2)
\end{equation}
\end{lemma}
Applying the lemma and Equation (\ref{eqn:pgmixture4}) to the case at hand, it is clear that
$$
\hat{\omega}^{(g)}_{ij} = \frac{n_{ij}}{\psi_{ij}} \tanh(\psi_{ij}/2) \, .
$$

We summarize the resulting EM algorithm in Figure \ref{algo:EM2x2.1}.  We also describe an ECM algorithm \citep{meng:rubin:1993} in Figure \ref{algo:EM2x2.2}, whereby $\mu$ and $\Sigma$ are also iteratively updated by conditional maximum likelihood, given the current estimate $\Psi^{(g)}$.  Two obvious hybrid strategyies, omitted from either figure, are: (1) to fix $\Sigma$, while still iteratively updating $\mu$; and (2) to further regularize $\mu$ and $\Sigma$ using a prior distribution.

\begin{figure}[t!]
\begin{center}
\fbox{
\begin{minipage}{32pc}
\begin{center}
\textbf{Algorithm 2: ECM for $2\times2\times N$ tables \\
(normal prior, unknown $\mu$ and $\Sigma$)}
\end{center}

Begin with an initial guess $\Psi^{(1)}, \mu^{(1)}, \Sigma^{(1)}$.

\begin{description}
\item[For] iteration $g = 1, 2, \ldots$

\begin{description}
\item[E Step:] For $i=1:n$ and $j=1:2$, set
$$
\omega_{ij}^{(g)} :=  \frac{n_{ij}}{\psi^{(g)}_{ij}} \tanh(\psi^{(g)}_{ij}/2) \, .
$$

\item[CM Step:] Update $\Psi$, $\mu$, and $\Sigma$ in turn.

\begin{description}

\item[Update $\Psi$:] For $i=1:n$ set
$$
\psi^{(g+1)} := \left(\Omega_i^{(g)} + \Sigma_{(g)}^{-1} \right)^{-1} \left(\kappa_i + \Sigma_{(g)}^{-1} \mu^{(g)} \right)
$$
for $\Omega_i^{(g)} = \mbox{diag}(\omega_{i1}^{(g)}, \omega_{i2}^{(g)} )$ and $\kappa_i = (y_{i1} - n_{i1}/2, y_{i2} - n_{i2}/2)'$.

\item[Update $\mu$ and $\Sigma$:]  Let 
\begin{eqnarray*}
\mu^{(g)} &=& N^{-1} \sum_{i=1}^N \psi_i^{(g)} \\
\Sigma^{(g)} &=& N^{-1} \sum_{i=1}^N (\psi_i^{(g)} - \mu^{(g)}) (\psi_i^{(g)} - \mu^{(g)})' \, .
\end{eqnarray*}

\end{description}

\end{description}

\item[End] when the sequence of estimates $\{\Psi^{(1)}, \Psi^{(2)}, \ldots \}$ has converged.

\end{description}

\end{minipage}
}
\vspace{0.5pc}
\caption{\label{algo:EM2x2.2} An ECM algorithm for estimating log-odds ratios in a $2\times2 \times N$ table where hyperparameters are estimated by maximum likelihood.}
\end{center}
\end{figure}

\section{Gibbs sampling}
\label{sec:PGGibbs}

To explore the joint posterior distribution over the log-odds ratios $\{\psi_{ij}\}$, we use a simple Gibbs-sampling algorithm.  The relevant conditional distributions for $\psi_{ij}$ and $\omega_{ij}$ are read off directly from Equations (\ref{eqn:pgmixture3}) and (\ref{eqn:pgmixture4}), and need no further elaboration.  We draw attention only to one curious property of these Gibbs updates: the conditional posterior distribution for $\omega_{ij}$ does not have an explicit closed-form density representation, but it is still very easy to sample from.  

We incorporate uncertainty in $(\mu,\Sigma)$ via a normal-Wishart hyperprior for $\mu$ and $\Lambda=\Sigma^{-1}$, leading to joint distributions of the form
\begin{align*}
p(\mu,\Lambda) &\propto |\Lambda|^{\frac{d-3}{2}} \exp \left ( -\frac{1}{2} \tr(B\Lambda) \right )\\
p(\psi,\mu,\Lambda) &\propto |\Lambda|^\frac{N}{2} 
 \exp \left (-\frac{1}{2} \sum_{i=1}^N (\psi^i-\mu)'\Lambda(\psi^i-\mu) \right ) \cdot |\Lambda|^{\frac{d-3}{2}} 
 \exp \left ( -\frac{1}{2} \tr(B\Lambda) \right ) \, .
\end{align*}
By comparison \citep{skene:wakefield:1990} used the improper uniform prior for $p(\mu)$, and a $\mathcal{IW}(d,B)$ prior for $\Sigma = \Lambda^{-1}$.  Applying standard theory of the conjugate normal--Wishart family, this models leads to conditionals of the form
\begin{align*}
( \mu \mid \Omega,\Psi, \Sigma ) &\sim \mathcal{N} \left ( N^{-1} \sum_{i=1}^N \psi^i, N^{-1} \Sigma \right )\\
( \Sigma \mid \Psi, \mu ) &\sim \mathcal{IW} \left (d+N, B+\sum_{i=1}^N(\psi^i-\mu)(\psi^i-\mu)' \right ) \, .
\end{align*}
The prior expectation of $\Sigma$ is given by
\begin{align*}
\E(\Sigma) = \E(\Lambda^{-1}) = \frac{B}{d-3}.
\end{align*}

The full MCMC is summarized in Figure \ref{algo:MCMC2x2}.  The only non-standard part of this algorithm is the generation of random variables from a Polya--Gamma distribution.  To do this, we use the representation in (\ref{eqn:PGdistribution1}), and truncate the sum of Gamma random variables after some large number $K$.  We have found that $K=200$ works well in practice, and that larger values made no discernible difference to the sum in the cases we examined.  Clearly this is an important (and easy) part of the sampler to check in examining the robustness of inferences.

The generation of so many Gamma random variables, merely to simulate a single Polya--Gamma random variable, may seem onerous.  But this can be done very rapidly on modern computers, and is far less time-consuming than it appears.  Moreover, multi-core processing environments are rapidly becoming the norm, and for tasks such as this offer a speedup that is essentially linear in the number of cores available.

\begin{figure}
\begin{center}
\fbox{
\begin{minipage}{32pc}
\begin{center}
\textbf{Algorithm 3: Exact Gibbs sampling for $2\times2\times N$ tables \\
(normal prior for $\psi_i$, N-IW prior for $\mu$ and $\Sigma$)}
\end{center}

Begin with an initial guess $\Psi^{(1)}, \mu^{(1)}, \Sigma^{(1)}$.

\begin{description}

\item[Update $\Omega$:] Draw each $\omega_{ij}$ as
$$
(\omega_{ij} \mid \psi_{ij}, D) \sim \PG\left( n_{ij}, \psi_{ij} \right) \, .
$$

\item[Update $\Psi$:] Draw each $\psi_{ij}$ as
\begin{eqnarray*}
(\psi_i \mid D, \Omega_i) &\sim& \N(m_i, V_{\Omega_i}) \, ,
\end{eqnarray*}
where
\begin{eqnarray*}
\Omega_i &=& \diag(\omega_{i1}, \omega_{i1}) \\
V_{\Omega_i}^{-1} &=& \Omega_i + \Sigma^{-1} \\
m_i &=& V_{\Omega_i} (\kappa_i + \Sigma^{-1} \mu) \\
\kappa_i &=& (y_{i1} - n_{i1}/2, y_{i2} - n_{i2}/2)' \, .
\end{eqnarray*}
\item[Update $\mu$ and $\Sigma$:]  Draw 
\begin{align*}
( \mu \mid \Psi, \Sigma ) &\sim \mathcal{N} \left ( N^{-1} \sum_{i=1}^N \psi_i, N^{-1} \Sigma \right )\\
( \Sigma \mid \Psi, \mu ) &\sim \mathcal{IW} \left (d+N, B+\sum_{i=1}^N(\psi_i-\mu)(\psi_i-\mu)' \right ) \, .
\end{align*}

\end{description}

\end{minipage}
}
\vspace{0.5pc}
\caption{\label{algo:MCMC2x2} A Gibbs-sampling algorithm for exploring the posterior distribution for the log-odds ratios in a $2\times2 \times N$ table.}
\end{center}
\end{figure}

\section{Example: a multi-center study on topical creams}
\label{sec:skenewakefieldexample}

To illustrate our Gibbs sampler, we analyze the data in Table \ref{tab:skenewakeexample}, from a multi-center study assessing the effectiveness of topical creams.  In contrast to the original analysis in \citet{skene:wakefield:1990}, we are able to avoid numerical integration by using the Gibbs sampler previously described (Algorithm 3).

\begin{figure}
\begin{center}
\includegraphics[width=5.5in]{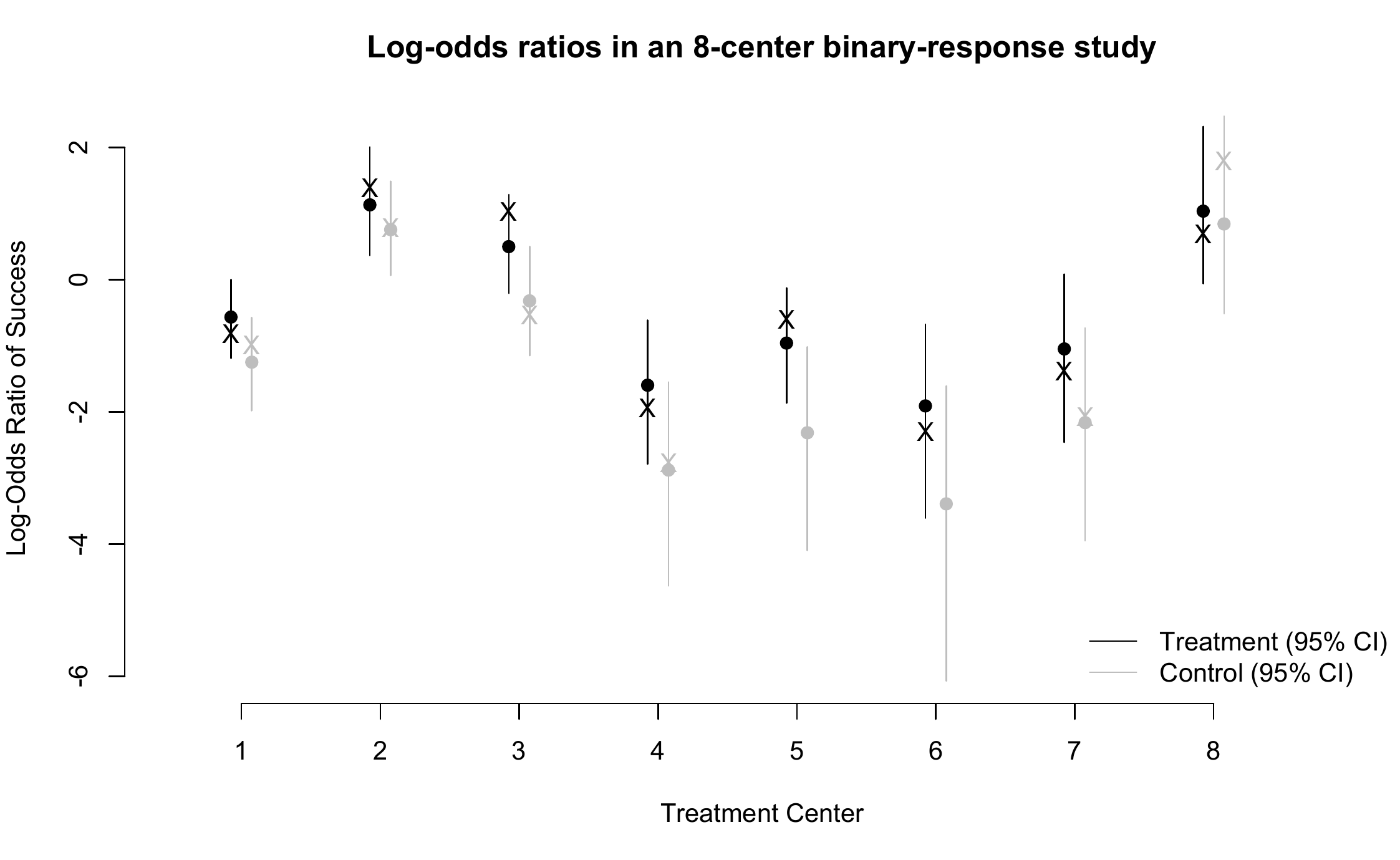}\\
\includegraphics[width=3.5in]{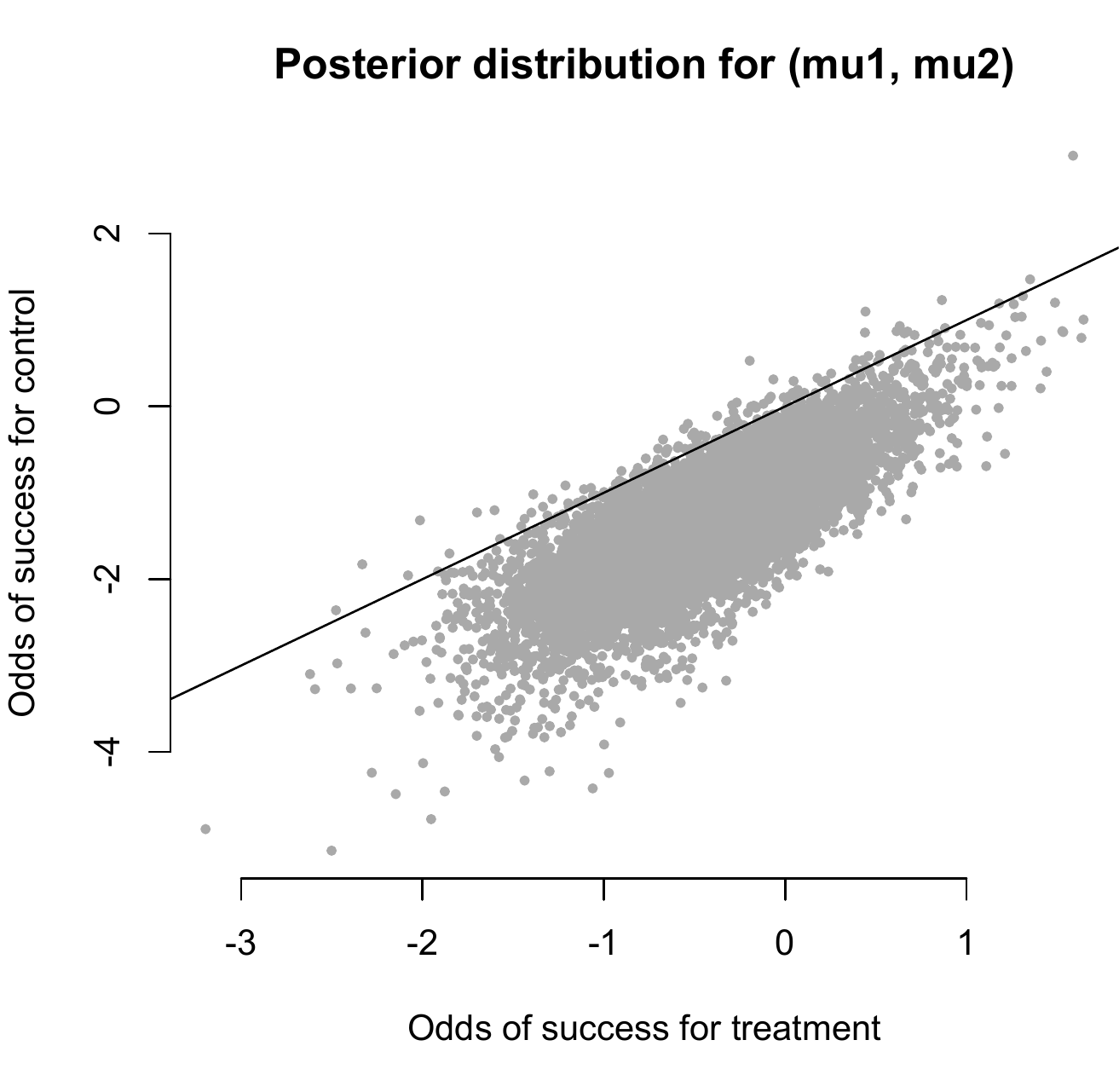} 
\caption{\label{fig:skenewakedata} Top: Posterior distributions for the log-odds ratio for each of the 8 centers in the topical-cream study from \citet{skene:wakefield:1990}.  The vertical lines are central $95\%$ posterior credible intervals; the dots are the posterior means; and the X's are the maximum-likelihood estimates of the log-odds ratios, with no shrinkage among the treatment centers.  Note that the MLE is $\psi_{i2} = -\infty$ for the control group in centers 5 and 6, as no successes were observed.  Bottom: draws from the joint posterior for $\mu = (\mu_1, \mu_2)'$, with the black line indicating the line where the two means are equal. }
\end{center}
\end{figure}

We use a normal-Wishart prior, as described above.   Hyperparameters were chosen to match Table II from \citet{skene:wakefield:1990}, who parameterize the model in terms of the prior expected values for $\rho$, $\sigma^2_{\delta}$, and $\sigma^2_{\lambda}$, where
$$
\Sigma = \left(
\begin{array}{cc}
\sigma^2_{\delta} & \rho \\
\rho & \sigma^2_{\lambda}
\end{array}
\right) \, .
$$
To match their choices, we use the following identity that codifies a relationship between the hyperparameters $B$ and $d$, and the prior moments for marginal variances and the correlation coefficient.  If $\Sigma \sim \mathcal{IW}(d, B)$, then
\begin{align*}
B = (d-3) \left[ \begin{array}{cc} \E(\sigma_\lambda^2)+\E(\sigma_\delta^2)+2\E(\rho) \sqrt{\E(\sigma_\lambda^2)\E(\sigma_\delta^2)} & \E(\sigma_\lambda^2) + \E(\rho) \sqrt{\E(\sigma_\lambda^2)\E(\sigma_\delta^2)} \\ \E(\sigma_\lambda^2) + \E(\rho) \sqrt{\E(\sigma_\lambda^2)\E(\sigma_\delta^2)} & \E(\sigma_\lambda^2) \end{array} \right] \, .
\end{align*}
In this way we are able to map from pre-specified moments to hyperparameters, ending up with $d = 4$ and
$$
B = \left(
\begin{array}{cc}
0.754 & 0.857 \\
0.857 & 1.480
\end{array}
\right) \, .
$$

The results of fitting this model via Gibbs sampling are summarized in Figure \ref{fig:skenewakedata}, which compares the posterior distribution of $\Psi$ with the maximum-likelihood estimator (which allows no pooling of information across the treatment centers).  Observe the effect of the shrinkage induced by the Bayesian model, particularly in the case of centers 5 and 6.  Also note that---while no individual center seems to produce overwhelming evidence that the treatment improves upon the control---the posterior distribution for $\mu$ supports the efficacy of the treatment quite strongly.  This shows the potential of our method for quantifying uncertainty in meta-analyses precisely of this kind.

\section{Prior choice}
\label{sec:priorchoice}

\subsection{Hierarchical Z priors}

Until now we have used a normal prior for each two-vector $\psi_i$, one which incorporates subjective information via $\mu$ and $\Sigma$.  For many applications this will involve no difficulty.  But as a default procedure, it poses two issues.  First, the tails of the normal prior are thinner than the tails of the logistic likelihood, a situation widely known to yield non-robust inferences \citep[e.g.][]{west:1984}.  This issue is particularly acute in the case of the logit likelihood, which is highly sensitive to large values.  Second, in the spirit of \citet{gelman:etal:2008}, we would like a ``default'' or weakly informative prior for log-odds ratios that lies somewhere between two extremes: fully informative priors, and formal noninformative priors (e.g.~reference priors).  This is particularly crucial if one intends to use the framework for model selection, in which noninformative priors cannot in general be used.

In choosing such a default prior, it is crucial to get two things right: the tails, and the scale.  To this end, \citet{gelman:etal:2008} propose the use of a Student-$t$ prior in logistic regression, where the likelihood can be well approximated by a $t_7$ with a scale parameter of $2.5$.  We agree with their basic reasoning leading to this choice.  Our only modification of their framework is to show how, using our data-augmentation scheme, one may conduct exact inference using a logistic-Z prior rather than a $t$ prior.  This will directly (rather than approximately) match the tails of the prior with that of the likelihood.  Moreover, it will do so with no extra model complexity or computational cost, compared with the Student-$t$ case.

To show this, we revisit some basic distributional properties of log-odds ratios.  Let $p$ be a success probability
and $ \psi $ the corresponding log-odds ratio. 
By definition, $ \psi \stackrel{D}{=} \log \left ( p/(1-p) \right ) $
with inverse  $ p = e^\psi / ( 1 + e^\psi ) $. The Jacobian
is $ \partial \psi / \partial p = ( 1 -p)^{-2} $. This leads to the following distributional identity:
$$ 
p \sim Be ( a , b ) \; {\rm implies} \; 
\psi = \log \left ( \frac{p}{1-p} \right ) \sim Z \left ( a,b,1,0 \right ) \, ,
$$
where $Z$ denotes Fisher's Z distribution \citep{fisher:1921,bn:kent:sorensen:1982}.  For example, Jeffreys' prior for a proportion $p$ is
$$
p ( p) = \frac{1}{\pi} \frac{1}{ \sqrt{p ( 1-p)}} \, ,
$$
which implies that
$$
p( \psi) = \frac{1}{\pi} \frac{ e^{ \frac{1}{2} \psi} }{ 1+ e^\psi } \, ,
$$
a $ Z \left ( \frac{1}{2} , \frac{1}{2} , 1 , 0 \right ) $ prior for the log-odds.

One possible choice of (independent) prior is therefore just a product of $Z(a_{ij}, b_{ij}, 1, 0)$ distributions for the $\psi_{ij}$'s, leading to a posterior of the form
\begin{eqnarray*}
p(\Psi \mid D) &\propto& \prod_{i=1}^N \left\{ p(\psi_1) p(D_1 \mid \psi_1) \ p(\psi_2) p(D_2 \mid \psi_2) \right\} \\
&\propto& \prod_{i=1}^N \left\{ \frac{  e^{ y_{i1} \psi_{i1} } }{ ( 1 + e^{\psi_{i1}} )^{n_{i1}} }
 \frac{  e^{ y_{i2} \psi_{i2} } }{ ( 1 + e^{\psi_{i2}} )^{n_{i2}} }
 \frac{  e^{ a_{i1} \psi_{i1} } }{ ( 1 + e^{\psi_{i1}} )^{a_{i1} + b_{i1} } }
 \frac{  e^{ a_{i2} \psi_{i2} } }{ ( 1 + e^{\psi_{i2}} )^{a_{i2} + b_{i2} } }  \right\} \\
		 &\propto& \prod_{i=1}^N \left\{ \frac{  e^{ (y_{i1} + a_{i1}) \psi_{i1} } }{ ( 1 + e^{\psi_{i1}} )^{n_{i1} + a_{i1} + b_{i1} } }
 \frac{  e^{ (y_{i2} + a_{i2}) \psi_{i2} } }{ ( 1 + e^{\psi_{i2}} )^{n_{i2} + a_{i2} + b_{i2} } }   \right\} \, .
\end{eqnarray*}
This leads to an obvious modification of the closed-form updates in the EM and Gibbs-sampling algorithms already presented, and induces no extra computational difficulty.  Under this framework, $a_{ij}$ and $b_{ij}$ can be interpreted as pseudo-data---specifically, the prior number of successes and failures at center $i$ for treatment $j$. A reasonable default choice might be $a_{ij} = b_{ij} = 1/2$, following Jeffreys' original argument.

This choice, however, precludes the possibility of learning about hyperparameters, and therefore pooling information across tables.  To allow this within the context of a default prior specification, we propose a hierarchical $Z$ prior with fixed scale, with tails that will match the likelihood.  Following \citet{bn:kent:sorensen:1982}, this can be represented as a variance-mean mixture of bivariate normals with respect to a Polya mixing distribution:
\begin{eqnarray*}
(\psi_i \mid \mu, \omega_0) &\sim& \N(\omega_0 \mu, \omega_0 C) \\
\omega_0 &\sim& \Pol(1/2, 1/2) \\
\mu &\sim& \N(0, I) \, ,
\end{eqnarray*}
where $C$ is some fixed correlation matrix (perhaps the identity).  This entails only minimal changes to the Gibbs-sampling updates for $\psi_i$ and $\mu$ in Algorithm 3.  The only additional step is the simulation of a single exponentially tilted Polya random variable, which can be done using the methods described in \citet{gramacy:polson:2010}.

\subsection{Other shrinkage priors}

Our hierarchical normal representation for the likelihood means that Bayesian versions of penalized-likelihood procedures can easily be used to yield regularized estimates of log-odds.  Consider, for example, the model where
\begin{eqnarray*}
(z_i \mid \Lambda_i, \mu, \Sigma) &\sim& \N(0, \Lambda_i) \\
\Lambda_i &=& \diag(\lambda_{i1}, \lambda_{i2}) \\
z_i &=& (\psi_{i1}, \psi_{i2})' \\
\lambda_{ij} &\sim& p(\lambda_{ij}) \, .
\end{eqnarray*}
If, for example, $\lambda_{ij} \sim \mbox{Ex}(2)$, then we have specified a lasso-type prior for $\psi_{i1}$, as well as for the contrast $\psi_{i1} - \psi_{i2}$.  Many other choices are possible (e.g.~horseshoe or bridge priors), with \citet{Polson:Scott:2010a} providing an extensive bibliography.  The posterior mode under such a specification can---if warranted by the data---collapse to a solution where $z_{i2} = 0$, in which case the treatment and control are estimated to be equally effective at treatment center $i$.

\section{Generalizations}
\label{sec:generalizations}

Now consider a multi-center, multinomial response study with more than two treatment arms.  This can be modeled using hierarchy of $N$ different two-way tables, each having the same $J$ treatment regimes and $K$ possible outcomes.  The data D consist of triply indexed outcomes $y_{ijk}$, each indicating the number of observations in center $i$ and treatment $j$ with outcome $k$.  We let $n_{ij} = \sum_k y_{ij}$ indicate the number of subjects assigned to have treatment $j$ at center $k$.

Let $P = \{p_{ijk}\}$ denote the set of probabilities that a subject in center $i$ with treatment $j$ experiences outcome $k$, such that $\sum_k p_{ijk} = 1$ for all $i, j$.  Given these probabilities, the full likelihood is
$$
L(P) = \prod_{i=1}^N \prod_{j=1}^J \prod_{k=1}^K p_{ijk}^{y_{ijk}} \, .
$$

Following \citet{leonard:1975}, we model these probabilities using a logistic transformation.  Let
$$
p_{ijk} = \frac{\exp(\psi_{ijk})}{\sum_{l=1}^K \exp(\psi_{ijl}) } \, .
$$
We assume an exchangeable matrix-normal prior at the level of treatment centers:
$$
\psi_i \sim \N(M, \Sigma_R, \Sigma_C) \, ,
$$
where $\psi_{i}$ is the matrix whose $(j,k)$ entry is $\psi_{ijk}$; $M$ is the mean matrix; and $\Sigma_R$ and $\Sigma_C$ are row- and column-specific covariance matrices, respectively. See \citet{dawid81} for further details on matrix-normal theory.  Note that, for identifiability, we set $\psi_{ijK} = 0$, implying that $\Sigma_C$ is of dimension $K-1$.

This leads to a posterior of the form
$$
p(\Psi \mid D) =  \cdot \prod_{i=1}^N \left[ p(\psi_i) \cdot \prod_{j=1}^J \prod_{k=1}^K  \left( \frac{\exp(\psi_{ijk})}{\sum_{l=1}^K \exp(\psi_{ijl}) } \right)^{y_{ijk}} \right] \, ,
$$
suppressing any dependence on $(M, \Sigma_R, \Sigma_C)$ for notational ease.

To show that this fits within the Polya--Gamma framework, we use a similar approach to \citet{holmes:held:2006}, rewriting each probability as
\begin{eqnarray*}
p_{ijk} &=& \frac{\exp(\psi_{ijk})}{\sum_{l \neq k} \exp(\psi_{ijl})  + \exp(\psi_{ijk})} \\
&=&  \frac{e^{\psi_{ijk}-c_{ijk}}} {1  + e^{\psi_{ijk}-c_{ijk}}} \, ,
\end{eqnarray*}
where $c_{ijk} = \log \{ \sum_{l \neq k} \exp(\psi_{ijl}) \}$ is implicitly a function of the other $\psi_{ijl}$'s for $l \neq k$.

We now fix values of $i$ and $k$ and examine the conditional posterior distribution for $\psi_{i \cdot k} = (\psi_{i1k}, \ldots, \psi_{iJk})'$, given $\psi_{i \cdot l}$ for $l \neq k$:
\begin{eqnarray*}
p( \psi_{i \cdot k} \mid D, \psi_{i \cdot (-k)}) &\propto& p(\psi_{i \cdot k} \mid \psi_{i \cdot (-k)}) \cdot  \prod_{j=1}^J  \left( \frac{e^{\psi_{ijk}-c_{ijk}}} {1  + e^{\psi_{ijk}-c_{ijk}}} \right)^{y_{ijk}} \left( \frac{1}{1  + e^{\psi_{ijk}-c_{ijk}}} \right)^{n_{ij} - y_{ijk}} \\
&=&  p(\psi_{i \cdot k} \mid \psi_{i \cdot (-k)}) \cdot  \prod_{j=1}^J  \frac{e^{y_{ijk}(\psi_{ijk}-c_{ijk})}} {(1  + e^{\psi_{ijk}-c_{ijk}})^{n_{ij}}}
\end{eqnarray*}

This is simply a multivariate version of the same bivariate form in (\ref{eqn:posterior2x2}).  Appealing to the theory of Polya--Gamma random variables outlined in the appendix, we may express this as:
\begin{eqnarray*}
p( \psi_{i \cdot k} \mid D, \psi_{i \cdot (-k)}) &\propto& p(\psi_{i \cdot k} \mid \psi_{i \cdot (-k)}) \cdot  \prod_{j=1}^J  
\frac{  e^ { \kappa_{ijk} [\psi_{ijk} - c_{ijk}] } } { \cosh^{n_{ij}} ([\psi_{ijk} - c_{ijk}]/2)} \\
&=&  p(\psi_{i \cdot k} \mid \psi_{i \cdot (-k)}) \cdot \prod_{j=1}^J \left[ e^{ \kappa_{ijk} [\psi_{ijk} - c_{ijk}]} \cdot 
\E \left\{ e^{-\omega_{ijk} [\psi_{ijk} - c_{ijk}]^2/2 } \right\} \right] \, ,
\end{eqnarray*}
where $\omega_{ijk} \sim \PG(n_{ij}, 0)$, $j=1,\ldots,J$; and $\kappa_{ijk} = y_{ijk} - n_{ij}/2$.  Given $\{ \omega_{i j k} \}$ for $j=1, \ldots, J$, all of these terms will combine in a single normal kernel, whose mean and covariance structure will depend heavily upon the particular choices of hyperparameters in the matrix-normal prior for $\psi_i$.  Each $\omega_{ijk}$ term, moreover has conditional posterior distribution
$$
(\omega_{ijk} \mid \psi_{ijk}) \sim \PG(n_{ij}, \psi_{ijk} - c_{ijk}) \, ,
$$
leading to a simple MCMC that loops over centers and responses, drawing each vector of parameters $\psi_{i \cdot k}$ (that is, for all treatments at once) conditional on the other $\psi_{i \cdot (-k)}$'s.

\section{Final remarks}
\label{sec:discussion}

We have shown that default Bayesian inference for multi-way categorical data can be implemented using a data augmentation scheme
based on Polya-Gamma distributions. This leads to simple Gibbs and EM algorithms for posterior computation that exploit standard normal linear-model theory.

It also opens the door for exact Bayesian treatments of many modern-day machine-learning classification methods based on mixtures of logits.  Indeed, many likelihood functions long thought to be intractable resemble the sum-of-exponentials form in the multinomial logit model of Section \ref{sec:generalizations}; two prominent examples are restricted Boltzmann machines \citep{salakhutdinov:etal:2007} and logistic-normal topic models \citep{blei:lafferty:2007}.  Applying the Polya--Gamma mixture framework to such problems is currently an active area of research.

A number of technical details of our latent-variable representation are worth further comment.  First, the dimensionality of the set of latent $\omega_{ij}$'s does not depend on the sample size $n_{ij}$ for each cell of the table.  Rather, the sample size only affects the distribution of these latent variables. Therefore, our EM and MCMC algorithms are more parsimonious than traditional approaches that require one latent variable for each observation.

Second, posterior updating via exponential tilting is a quite general situation that arises in Bayesian inference incorporating latent variables.  For example, the posterior distribution of $ \omega $ that arises under normal data with precision $\omega$ and a $\PG(a,0)$ prior 
is precisely an exponentially titled $\PG(a,0)$ random variable.  This led to our characterization of the general $\PG(a, c)$ class. 
  
Notice, moreover, that one may identify the conditional posterior for $\omega_{ij}$ 
strictly using its moment-generating function, without ever appealing to Bayes' rule for
density functions.  This follows the L\'evy-penalty framework of \citet{Polson:Scott:2010b}
and relates to work by \citet{ciesielski:taylor:1962}, who use a similar argument to 
characterize sojourn times of Brownian motion.  It offers the advantage of suggesting a simple route for simulating $\PG(a,c)$ random variables, 
a crucial step in our computational results.  Doubtless there are many other modeling situations where the basic idea is also applicable, or will lead to new insights.

\appendix

\section{Properties of Polya--Gamma random variables}

\subsection{The case $\PG(a,0)$}

We construct the family of Polya--Gamma random variables as follows.  Following \citet{devroye:2009}, a random variable $J$ has a Jacobi distribution if
\begin{equation}
\label{eqn:jacobidistribution1}
J \stackrel{D}{=} \frac{2}{\pi^2} \sum_{k=1}^{\infty} \frac{e_k}{(k-1/2)^2} \, ,
\end{equation}
where the $e_k$ are independent, standard exponential random variables.  The moment-generating function of this distribution is
$$
\E(e^{-t J}) = \frac{1}{\cosh(\sqrt{2t})} \, .
$$
The density of this distribution is expressible as a multi-scale mixture of inverse-Gaussians; all moments are finite and expressible in terms of Riemann zeta functions.  For details, see \citet{devroye:2009}.

The Jacobi is related to the Polya distribution \citep{bn:kent:sorensen:1982}, 
in that if $J$ has a Jacobi distribution, and $\omega \stackrel{D}{=} J/4$, then $\omega \sim \Pol(1/2, 1/2)$.

Let $\omega_k \sim  \Pol(1/2, 1/2)$ for $k = 1 ,\ldots, n$ be a set of independent 
Polya-distributed random variables. 
A $\PG(n,0)$ random variable is then defined by the sum $\omega_{\star} \stackrel{D}{=} \sum_{k=1}^n \omega_k$. 
Its moment generating function follows from that of  a Jacobi distribution, namely
$$
\E\{\exp(-\omega_k t)\} = \frac{1}{\cosh(\sqrt{t/2})} \; 
{\rm and} \; 
\E\{\exp(-\omega^{\star} t)\} = \frac{1}{\cosh^n(\sqrt{t/2} )} \, .
$$
The name ``Polya--Gamma'' arises from the following observation.  From (\ref{eqn:jacobidistribution1}),
$$
\omega_{\star} \stackrel{D}{=} \sum_{l=1}^{n} \left( \frac{1}{2 \pi^2} \sum_{k=1}^{\infty} \frac{e_{l,k}}{(k-1/2)^2} \right) \, ,
$$
where $e_{l,k}$ are independent exponential random variables.  Rearranging terms,
\begin{eqnarray*}
\omega_{\star} &\stackrel{D}{=}&  \frac{1}{2 \pi^2} \sum_{k=1}^{\infty} \frac{\sum_{l=1}^{n} e_{l,k}}{(k-1/2)^2}  \\
			 &\stackrel{D}{=}&  \frac{1}{2 \pi^2} \sum_{k=1}^{\infty} \frac{g_k }{(k-1/2)^2} \, ,
\end{eqnarray*}
where $g_k$ are i.i.d. Gamma$(n,1)$ random variables.  More generally we may replace $n$ with any positive real $a$.

\subsection{The general case}

The general $\PG(a,c)$ class arises through an exponential tilting of the $\PG(a,0)$ density:
\begin{equation}
\label{eqn:polyagamma.generalpdf}
p_{a,c}(\omega) = \frac{\exp \left( -\frac{c^2}{2} \omega \right) p_{a,0}(\omega)}{ \E_{a,0} \left\{ \exp(-\frac{c^2}{2} \omega) \right\} } \, ,
\end{equation}
where $ p_{a,0}(\omega)$ is the density of a $\PG(a,0)$ random variable.  Using the above results, along with Euler's expansion of the $\cosh$ function, write the moment-generating function of this distribution as
\begin{align*}
 \E_{a,c} \left\{ \exp\left( -\frac{1}{2} \omega t \right) \right\} 
  &=  \frac{\cosh^{a} \left( \frac{c}{2} \right)}  {\cosh^{a} \left( \frac{\sqrt{c^2+t}}{2} \right) } \\
& = \prod_{k=1}^\infty \left( \frac{1+\frac{c^2}{4(k-1/2)^2\pi^2}} {1+\frac{c^2+t}{4(k-1/2)^2\pi^2}} \right)^{a} \\
 & = \prod_{k=1}^\infty (1+ d_k^{-1} t )^{-a} \; {\rm where} \;  d_k = 4\left(k-\frac{1}{2}\right)^2\pi^2 + c^2 \; .
\end{align*}

We can therefore write a $\PG(a,c)$ random variable as
\begin{align*}
\omega \stackrel{D}{=} 2 \sum_{k=1}^\infty \frac{\Ga(a,1)}{d_k} = 
\frac{1}{2 \pi^2} \sum_{k=1}^\infty \frac{\Ga(a,1)}{(k-\frac{1}{2})^2  + c^2/(4 \pi^2) } \, ,
\end{align*}
appealing to the moment-generating function of the gamma distribution.

\section{Proofs of main results}

\subsection{Theorem \ref{thm:2x2PGmixture}}

With all the distributional theory of the previous section in place, the proofs of our main results will proceed very straightforwardly.

\begin{proof}
Use the expressions for the moment-generating function of a Polya--Gamma random variable (given above) to write the likelihood in (\ref{eqn:posterior2x2}) as
\begin{eqnarray*}
\frac{( e^{\psi_1})^{y_1} } {( 1 + e^{\psi_1})^{n_1} }   \cdot \frac{( e^{\psi_2})^{y_2} } {( 1 + e^{\psi_2})^{n_2} } 
&=&  \frac{  2^{-n_1} \exp \{ \kappa_1 \psi_1 \} } { \cosh^{n_1} (\psi_1/2)}
\cdot \frac{  2^{-n_2} \exp \{ \kappa_2 \psi_2 \} } { \cosh^{n_2} (\psi_2/2)} \\ \\
&=& 2^{-(n_1 + n_2)} e^{\kappa_1 \psi_1} \ e^{\kappa_2 \psi_2}
\E \{ \exp(-\omega_1 \psi_1^2/2 \}  \  \E \{ \exp(-\omega_2 \psi_2^2/2 \} \, ,
\end{eqnarray*}
where $\omega_j \sim \PG(n_j, 0)$, $j=1,2$; and where we recall that $\kappa_j = y_j - n_j/2$.

Given particular values of $\omega_1$ and $\omega_2$, we can write $p(\psi \mid D, \Omega)$ as
\begin{eqnarray*}
p(\psi \mid D, \Omega) &\propto& \exp( \kappa_1 \psi_1 - \omega_1 \psi_1^2 /2 ) \cdot \exp( \kappa_2 \psi_2 - \omega_2 \psi^2_2/2 ) \ p(\psi \mid \mu, \Sigma) \\ \\
&\propto& \exp \left\{ - \frac{\omega_1}{2} \left( \psi_1 - \kappa_1/\omega_1  \right)^2 \right\}  
\cdot \exp \left\{ - \frac{\omega_2}{2} \left( \psi_2 - \kappa_2/\omega_2  \right)^2 \right\}  \ p(\psi \mid \mu, \Sigma) \, .
\end{eqnarray*}
Since $p(\psi \mid \mu, \Sigma)$ is a bivariate normal prior, the posterior is conditionally normal, with the specific form given by Part A of the theorem.

Turning now to Part B, we observe that the conditional posterior $p(\omega_j \mid \psi_j, D)$ is of the same form as (\ref{eqn:polyagamma.generalpdf}), with $\psi_j = c$.  We therefore arrive at the result by straightforwardly applying the previous section's distributional theory for $\PG(a,c)$ random variables.

\end{proof}

\subsection{Lemma \ref{lem:PGmoment}}

\begin{proof}
From the moment generating function for $ \omega \sim \PG(a,0) $ density evaluated at $ \frac{1}{2} c^2 $ we have
\begin{align*}
\cosh^{-a} \left ( \frac{c}{2} \right ) & = \mathbb{E} \left (  e^{ - \frac{1}{2} \omega c^2 }   \right )\\
 &=  \int_0^\infty  e^{ - \frac{1}{2} \omega c^2 }    p_{\PG(a,0)} ( \omega ) d \omega \, .
\end{align*}
Taking logs and differentiating under the integral sign with respect to $c$ then gives the moment identity
$$
\mathbb{E} \left ( \omega  \right ) = \frac{1}{c} \frac{\partial}{\partial c} \log \cosh^{a} \left ( \frac{c}{2} \right ) \, .
$$
Simple algebra reduces this down to the form given earlier, 
$$
\mathbb{E} \left ( \omega  \right ) = \frac{a}{c} \tanh \left ( \frac{c}{2} \right ) \, .
$$
\end{proof}

\singlespace
\begin{footnotesize}
\bibliographystyle{abbrvnat}
\bibliography{masterbib}
\end{footnotesize}

\end{spacing}

\end{document}